\documentstyle[eqsecnum,prb,aps]{revtex}
\begin{document}
\draft
\columnsep -.375in
\twocolumn[
\title{
Fluctuations and Instabilities of Ferromagnetic Domain Wall
Pairs in an External Magnetic Field
}
\author{Hans-Benjamin Braun\cite{address}}
\address{
Department of Physics, University of California San Diego, La
Jolla CA 92093-0319
}
\date{May 9, 1994}
\maketitle
\begin{abstract}
\widetext
A classical continuum model of
an effectively one-dimensional ferromagnet with exchange
and anisotropies of
hard and easy-axis type is considered.
If an external field is applied along the easy axis,
the lowest lying
topological excitations are shown to be
untwisted or twisted pairs of $\pi$-domain walls.
The fluctuations around these structures are investigated.
It is shown  that the fluctuations around the twisted and
untwisted domain wall pair are governed by the
same set of operators.  The untwisted domain wall pair has
exactly one unstable mode and thus represents
a critical nucleus for magnetization reversal in effectively
one-dimensional systems. The twisted domain wall pair is
stable for small external fields but becomes
unstable for large magnetic fields.
The former effect is related to thermally induced
coercivity reduction in elongated particles while
the latter effect is related to
``chopping" of twisted Bloch wall pairs in  thin films.
 In view of a statistical mechanical theory of magnetization
reversal which will be presented in a separate article,
the scattering phase shifts of spin waves around these
structures are calculated.
The applicability of the present theory
to magnetic thin films is discussed. Finally, it is noted
that the static properties of the present  model are
equivalent to those of a nonlinear $\sigma$-model
with anisotropies and an external field.
\end{abstract}
%
]
\narrowtext
\section{Introduction}

Macroscopic ferromagnetic samples \cite{Enzetal,sloncz}
consist of many domains
in which the magnetization is uniform and directed
along one of the minima of the crystalline
anisotropy. The phase boundaries  between such
regions are formed by domain walls (Bloch walls)
in which the magnetization vector
rotates continuously
between different anisotropy minima.
The formation of these domains is due to the long-range magnetostatic
forces which tend to avoid the formation of magnetostatic
charges at the sample surface. However, domain walls have
locally  planar symmetry and can therefore locally be described by
an effectively one-dimensional model \cite{Enzetal,sloncz}.

A one-dimensional description is also adequate for
elongated samples of mesoscopic size if the lateral
sample extension is less than a domain wall width.
Such particles are widely used in magnetic recording media,
e.g. $Cr O_2$ -particles \cite{koester} are amost perfect needles
with aspect ratios of up to 20:1. For this
reason and in view of tremendous recent progress in sample
preparation on the nanometer scale,
it is therefore of particular importance to study the model of an
effectively one dimensional ferromagnet in detail.

In the following we shall focus on a  description  of  the
magnetization within a classical field theory.
Such a formulation also provides
the starting point for a quantum mechanical theory
in the semiclassical limit \cite{Raj}.
The magnetization is treated as a classical vector of constant
magnitude and adjacent moments interact via exchange
thus giving rise to a ``stiffness" of the spin chain.
The present model contains single ion anisotropies
of hard- and easy-axis type which may have demagnetizing or
crystalline origin. In addition it includes an external field directed
along the  easy-axis.  Without an external field, this model
is also known as a ``biaxial ferromagnet".

The present model  has also been used to describe weakly coupled
one-dimensional (1D) ferromagnetic
chains \cite{hoog}. In effectively 1D antiferromagnets such as
TMMC  \cite{steinmik}, it emerges as an
effective model for the sublattice magnetization.
The dynamic version of this model without external field
and damping has been shown to be integrable \cite{sklyanin}
and reveals a surprisingly rich palette of soliton and breather
\cite{braunbrod} solutions, the solitons playing the role of
domain walls.

The simplest static, topological excitation in a biaxial ferromagnet
in the absence of an external field is the $\pi$-Bloch wall
\cite{bloch,landlif} (see Fig.\ref{blochwall}) which
constitutes the transition region between two equivalent
anisotropy minima. Its stability has been investigated by
Winter \cite{Winter} who explicitly derived
spin wave excitations. He showed that within the 1D system,
Bloch walls are stable save for the zero energy mode
which describes a rigid  translation of the domain wall.
Later Janak \cite{Janak} quantized the spin wave excitations
around a pinned domain wall and included demagnetizing effects
of spin waves running parallel to the domain wall.
Hornreich and Thomas \cite{hornthom} considered a biaxial
ferromagnet with an external field perpendicular to the easy axis.
They studied the instability of domain wall structures for large
external fields and gave variational stability boundaries
including demagnetizing effects of fluctuations.

In this paper we consider the different situation of an external
field applied along the easy axis without the  limitation to
large external fields. The external field removes
the degeneracy between the two
anisotropy minima and consequently only pairs of Bloch walls
can exist as  static solutions. The basic topological excitations
of this system are thus twisted and untwisted pairs of
$\pi$-Bloch walls.

Experiments \cite{niedoba} and numerical simulations
\cite{humphrey} suggest that the annihilation of
twisted domain-wall pairs in thin films
requires much larger external fields than that of
untwisted domain-wall pairs. Furthermore, the
observed \cite{koester} coercivity reduction
in elongated particles at finite temperatures has no
theoretical explanation.

In this work, it is shown that both of these effects are
related to the stability properties of twisted and
untwisted domain wall pairs. The primary aim of the
present paper is therefore a careful investigation
of the fluctuations around these structures.
We shall reveal the surprising fact that
fluctuations around the twisted and
untwisted domain wall pairs are described by the
same set of operators. This puts the stability
discussion of the untwisted and the twisted
domain wall pair on an equal footing.
It then follows immediately that the untwisted
domain wall pair has exactly one unstable mode
corresponding to an expansion or a shrinking of the structure.
The untwisted domain wall pair is thus identified as
a ``nucleus" of critical size in a first order phase transition
and thus plays a crucial role in thermally activated
magnetization reversal \cite{prlnucl}
in elongated particles. A detailed statistical
mechanical theory of magnetization reversal
will be presented in a forthcoming  paper \cite{braunnucl}.

Another immediate consequence of this relation is
the instability of the twisted domain wall pair
(or ``$2\pi$-Bloch wall") for large external fields
as has been discovered by Magyari  and Thomas
\cite{thomasmag} and independently in Ref. \onlinecite{diss}.
By a careful examination of the
nonlocal demagnetizing fields which are not included
in the model of a biaxial ferromagnet,  it is shown that
this effect should be observable in thin films. In particular, the
minimal attainable distance of two domain walls
is shown to decrease with increasing hard-axis ansiotropy.
It is emphasized that this effect is beyond the
otherwise highly successful description of
domain walls within Slonczewski's effective
model \cite{sloncz}. The present results  are also
crucial for the current design of vertical Bloch line
memories \cite{katti} whose read operations
rely \cite{humphrey} on a distinction
between domain wall pairs with different relative
sense of twist.

The work is organized as follows.
In section II we present the model and discuss its role
as an effective model which describes planar structures
in a  3D model {\it including} demagnetizing effects.
In section III  untwisted and twisted domain wall pairs are
presented and their energy is evaluated. It is shown that both
structures can be viewed as a coherent superposition of two
$\pi$-Bloch walls.
In section IV the operators governing the fluctuations
around the $2\pi$-Bloch wall and nucleus are derived.
In section V these results are applied to discuss the instabilities
of these structures.
In section VI, bound state energies and
scattering phase shifts of the fluctuation
operators are discussed analytically and
numerically  in view of a calculation of nucleation
rates of domain wall pairs.
The discussion of scattering phase shifts provides a lucid example
of the widely unknown version of Levinson's theorem in 1D:
Scattering phase shifts do not converge uniformly
to those of the operators that are obtained in the
limit of  small and large external fields.
In  section VII we show that the present model
can account for several different experimental configurations
in thin films and we shall show that  the nonlocal influence
of demagnetizing fields on twisted domain wall pairs
can be neglected for sufficiently  thin films.
The present model is thus
adequate for small domain wall distances in
sufficiently thin films where nonlocal
demagnetizing effects are shown to be negligible.

It is not necessary that the reader follows all details
of the present paper.  Those who are interested in
experimental implications may skip the more
formal sections IV and VI and
directly proceed to  section VII.

\section{The model}

In this we work we consider effectively one-dimensional
magnetization configurations described by the
following energy per unit area
\begin{eqnarray}
{\cal E}=\int dz \left\{{A\over M_0^2} [(\partial_z M_x)^2+
(\partial_z M_y)^2+(\partial_z M_z)^2] \right.\nonumber\\
\left.+{K_h\over M_0^2} M_z^2-{K_e\over M_0^2} M_x^2
-\! H_{\rm ext}
M_x\right\},
\label{e0}
\end{eqnarray}
where ${\bf M}={\bf M}(z)$, $\partial_z\equiv \partial/
\partial z$
and  $M_0 \equiv |{\bf M}|$ is the constant magnitude of the
magnetization.  The first term in (\ref{e0}) is the classical
counterpart of exchange energy and $A$ is
an exchange constant. The second term describes a hard-axis
anisotropy characterized by the anisotropy
constant $K_h>0$ thus rendering the $xy$-plane an
easy-plane. The rotational invariance in this easy-plane
is broken by an additional
easy-axis anisotropy with anisotropy constant $K_e>0$.
The last term in the integrand of (\ref{e0})
is the Zeeman term which is due to
an external field ${\bf H}_{\rm ext}$ pointing
along the easy-axis.

Apart from the description of the
(sublattice-)spin configuration
in 1D (anti-) ferromagnetic systems \cite{hoog,steinmik},
the energy (\ref{e0})
has found wide
applications \cite{Enzetal,sloncz} in the  description
of planar domain  walls and their mobilities in
bulk ferromagnets. As will be discussed in section VII,
it is also adequate for the description of domain wall pairs in
thin films.  Due to the absence of discussion in the
recent literature,  it seems convenient to review
how the energy (\ref{e0}) may be derived from the
energy  of arbitrary 3D-magnetization configurations
${\bf M}={\bf M}({\bf r})$ in a volume $V$
with inclusion of demagnetizing effects:
\begin{eqnarray}
E=\int_V d^3r \left\{ {A\over M_0^2} [(\nabla M_x)^2+
(\nabla M_y)^2+(\nabla M_z)^2]\right.\nonumber\\
\left.  -{  K_{e,{\rm cryst}}\over M_0^2   }   M_x^2  +
{   K_{h,{\rm cryst}}\over M_0^2  } M_z^2-\!
{1\over 2} {\bf H}_m \cdot{\bf M} -\! H_{\rm ext} M_x\right\}.
\label{e3d}
\end{eqnarray}
In contrast to (\ref{e0}), the first term  in the
integrand is the exchange term
in three dimensions while the second and third terms
describe crystalline easy- and hard-axis anisotropies
of strengths $K_{e,{\rm cryst}}, K_{h,{\rm cryst}}>0$,
respectively.
The fourth term is the demagnetizing energy with the
demagnetizing field ${\bf H}_m$ obeying the
magnetostatic Maxwell equations
$\nabla\times{\bf H}_m=0$, $\nabla\cdot{\bf B}=0$,
$({\bf B}={\bf H}_m+ 4\pi {\bf M})$.
They can be rewritten in the form of a Poisson equation
$\nabla^2 \Phi_m=4\pi\nabla\cdot {\bf M}$ with
the magnetostatic potential $\Phi_m$ defined
via ${\bf H}_m=-\nabla\Phi_m$.
The Poisson equation is integrated in a standard way,
and after splitting volume and surface terms we obtain
\begin{eqnarray}
{\bf H}_m({\bf r})=&-&\int _V d^3r' \;\rho_m({\bf r}')\;
{{\bf r}-{\bf r}'\over |{\bf r}-{\bf r}'|^3}
\nonumber\\
&+&\int_{\partial V} dS'\;\sigma_m({\bf r}') \;
{{\bf r}-{\bf r}'\over |{\bf r}-{\bf r}'|^3},
\label{Hm2}
\end{eqnarray}
where $\sigma_m(
{\bf r})\equiv {\bf M} ({\bf r})\cdot{\bf n}({\bf r})$ is the magnetic
surface charge (${\bf n}$ is the normal of the surface $\partial V$)
and $\rho_m({\bf r})=\nabla\cdot{\bf M}({\bf r})$
is the magnetic volume charge. Inserting
({\ref{Hm2}) into (\ref{e3d}) one recognizes that the
evaluation of a magnetization configuration ${\bf M} ({\bf r})$
by minimization of (\ref{e3d}) for given boundary conditions
is in general a hopeless task.

However, experiments reveal that the magnetization
distribution in the vicinity of a domain
wall in the bulk of a sample is a locally planar
structure. This suggests the existence of an effective energy
density which is of the form (\ref{e0}). In fact
restricting ourselves to planar structures ${\bf M}={\bf M}(z)$
and neglecting magnetic surface charges \cite{magcharge}
in (\ref{Hm2}) we obtain for an  infinite sample a
demagnetizing field of the form
\begin{equation}
{\bf H}_m(z)=-4\pi M_z(z) {\bf e}_z,
\label{Hm1d}
\end{equation}
where ${\bf e}_z$ is the unit vector in $z$-direction. For the
derivation of (\ref{Hm1d}) we have also assumed that
$M_z(\pm\infty)=0$.
After insertion of (\ref{Hm1d}) into (\ref{e3d}),
the demagnetizing energy takes the form
of a hard-axis anisotropy along the $z$-direction.
The underlying physical
picture is simple: A planar arrangement of parallel
dipoles has higher energy when the dipoles stick out
of the  plane than if they are in the plane.
The form of the  demagnetizing field (\ref{Hm1d}) is used to
analyze wall motion experiments in garnet films
\cite{Enzetal}.  For structures of planar
symmetry we thus may reduce (\ref{e3d}) to (\ref{e0}) provided
that
\begin{eqnarray}
K_h&=&K_{h,{\rm cryst}}+2\pi M_0^2>0,\\
K_e&=&K_{e,{\rm cryst}}>0.
\end{eqnarray}
This holds for a configuration as e.g. shown in Fig. \ref{thinfilms} a).
For other sample geometries and anisotropy configurations,
we can similarly express the effective  anisotropy constants
$K_e$, $K_h$ in (\ref{e0}) by shape and crystalline anisotropies.

To incorporate the constraint
${\bf M}^2=M_0^2=const.$ in Eq. (\ref{e0}), we use spherical
 coordinates defined by $\;{\bf M}/M_0$ $=(\sin\theta\cos\phi$,
$\sin\theta\sin\phi$, $\cos\theta)\;$.
Further it is convenient to introduce dimensionless
quantities by taking
the scales of length and energy per area as
\begin{equation}
 [x]=[y]=[z]=\sqrt{A\over K_e}
,\qquad [{\cal E}]=2\sqrt{AK_e}.
\label{dimless}
\end{equation}
Consequently, the units of the magnetic field are given by
$[H]=\sqrt{2 K_e}$.  The length $\sqrt{A/K_e}$ is the
width of the static $\pi$-Bloch wall,
and $2\sqrt{A K_e}$ is half the energy per unit area of the static
$\pi$-Bloch wall.  With these definitions, the
energy (\ref{e0}) becomes
\begin{eqnarray}
{\cal E}=\int_{-\infty}^\infty\!\! dz \left\{ {1\over 2}
[(\partial_z\theta)^2+\sin^2\theta
(\partial_z \phi)^2]\right. \nonumber\\
\left. -{1\over2}[\sin^2\theta\cos^2\phi\!-\!1]
+{Q^{-1}\over 2}\cos^2\theta\!-\! h\sin\theta\cos\phi\right\},
\label{e}
\end{eqnarray}
where $\theta=\theta(z)$ and $\phi=\phi(z)$.
The normalization is chosen such that the uniform states $\theta=\pi/2$
and $\phi=0$ or $\phi=\pi$ have zero
energy in the absence of an external field. In (\ref{e})
we have introduced the dimensionless anisotropy ratio
$Q>0$ with
\begin{equation}
Q={K_e\over K_h},
\label{Q}
\end{equation}
describing the ratio of easy- and hard-axis anisotropy in the effective
model (\ref{e0}). Note that this
is a slight extension of the common definition where $K_h=2\pi
M_0^2$. In (\ref{e}) we have also used the reduced external
field $h$ which is related  to the external field $H_{\rm ext}$
in laboratory units by
\begin{equation}
 h={H_{\rm ext} M_0\over 2 K_e}>0.
\label{h}
\end{equation}

At first sight, the choice of the coordinate frame in (\ref{e0}) and  (\ref{e})
might be surprising  since the polar angle is not measured relative to
the external field.
The advantage of such an
orientation is that the linearization in the angles $\theta$ and $\phi$
around structures confined to the $xy$-plane
is equivalent to a linearization in  a cartesian frame that is rotated
along this structure \cite{Winter,Janak}
but is simpler in practice.
Measuring $\theta$ from the external field would not allow linearization
in the azimuthal angle $\phi$
to describe spin wave excitations of a uniform state parallel to the
external field.

\section {Domain Wall Structures}

In the following  we shall focus on static easy-plane structures
of the model (\ref{e}).  It is shown  that
the only solitary easy-plane structures are twisted
and untwisted  pairs of  $\pi$-Bloch walls.
Simple representations are presented that
relate these  solutions to each other.

Inspecting (\ref{e}), we recognize that the hard-axis
anisotropy is minimized for $\theta=\pi/2$.
The corresponding
static structures then identically satisfy the Euler-Lagrange
equation $\delta{\cal E}/\delta\theta=0$  while the
Euler-Lagrange equation in $\phi$ reads
\begin{equation}
-{d^2\phi\over d z^2} + \sin\phi\cos\phi +h\sin\phi =0.
\label{EL}
\end{equation}
Upon integration with $d\phi/ dz$ we obtain the first integral
\begin{equation}
{1\over2}\left({d\phi\over dz}\right)^2 +V(\phi) =C,
\label{econs}
\end{equation}
with
\begin{equation}
V(\phi)={1\over 2}\cos^2\phi + h\cos\phi.
\label{epot}
\end{equation}
Eq. (\ref{econs}) has the form of an energy conservation for
a fictitious particle moving in the one dimensional potential $V(\phi)$.
By this analogy,  we can gain an overview \cite{diss} of all static
easy-plane structures. Note that the potential $V(\phi)$ is the
negative of anisotropy and external field
contributions  to  ${\cal E}$ for $\theta=\pi/2$ up
to an irrelevant constant.

Solitary solutions are now obtained as trajectories
of the fictitious particle starting from a
local maximum of $V(\phi)$. Due to the ``energy
conservation" (\ref{econs}) it will either creep into
a different maximum of the same height or, if it
started from a lower maximum, it will bounce back
into the same state.
For $h\neq 0$ the degeneracy between the maxima of $V$
(i.e., minima of ${\cal E}$)
is lifted and two distinct trajectories emerge. One trajectory
connects a global maximum of $V$ at $\phi=0$ with an
adjacent one at $\phi=\pm2\pi$. This
trajectory corresponds to a twisted pair of $\pi$-Bloch walls.
The second possible trajectory represents a localized
excursion from the lower maximum of $V$ at $\phi=\pm\pi$
which corresponds to an untwisted pair of  $\pi$-Bloch walls.
For other values of $C$ in (\ref{econs}), periodic
solutions \cite{diss} occur
which may be regarded as generalizations of the above solutions
to finite sample lengths.

Thus we have gained an overview over all possible solutions
without having solved the differential equation (\ref{econs}) in detail.
This analogue should also prove useful for different models
with other forms of the
anisotropy and different orientations of the external field.

Apart from the trivial symmetry arising from the representation
of ${\bf M}$ in terms of spherical coordinates, Eq.(\ref{EL}) is
invariant under the following symmetry operations
\begin{equation}
{\cal R}_x(\pi):\phi\mapsto -\phi,
\label{Rx}
\end{equation}
and
\begin{eqnarray}
{\cal T}:&&{\bf M}\mapsto -{\bf M},  \nonumber\\
    &&h\mapsto -h.
\label{S}
\end{eqnarray}
${\cal R}_x(\pi)$ corresponds to an (internal) rotation of the magnetization
by an angle $\pi$ around the $x$-axis ($\theta=\pi/2$),
whereas ${\cal T}$ represents
a time inversion. Therefore, all specific solutions quoted below
have equivalents arising through the action of ${\cal R}_x$, ${\cal T}$ and
${\cal T}\circ{\cal R}_x$.
For a given direction of the external field there are thus exactly two
equivalent structures related to each other by the action of
${\cal R}_x(\pi)$.
To classify the solutions it is also
convenient to introduce the twist
\begin{equation}
q(\phi)={1\over 2\pi} \int_{-\infty}^{\infty} dz\,{d\phi\over dz}.
\label{q}
\end{equation}
Single $\pi$-domain walls belong to $|q|=1/2$, whereas twisted
domain wall pairs have $|q|=1$ and untwisted pairs have $q=0$.
Note that ${\cal R}_x(\pi)$ changes the sign of the twist $q$,
whereas ${\cal T}$ leaves the twist invariant but reverses the
magnetization at infinity.

For a vanishing external field, $h=0$, (\ref{econs}) with
(\ref{epot}) may easily be integrated with the boundary
conditions $\partial_z\phi(\pm\infty)=0$,
$\phi(-\infty)=\pi$, and $\phi(\infty)=0$ to yield the $\pi$-Bloch wall
\begin{eqnarray}
\phi_K(z)&=&2\arctan e^{-z},\nonumber\\
\theta_K&=&\pi/2.
\label{kink}
\end{eqnarray}
The configuration (\ref{kink}) is shown in Fig. \ref{blochwall}.
The Bloch wall  represents a smooth transition
region between the two degenerate uniform states of minimal
anisotropy energy while
the magnetization always lies in the easy plane.
In (\ref{kink}) an  integration constant describing the
arbitrary wall position has been fixed such  that  the
$\pi$-Bloch wall is centered around the origin.
However,  as we shall see in Sec. IV, this degeneracy
with respect to translations will lead to a (Goldstone-)
mode of zero energy in the excitation spectrum.
The finite domain wall width arises through the balance of
exchange energy and uniaxial anisotropy, the former tending to
enlarge the transition region, the latter tending to narrow the
Bloch wall.

Inserting (\ref{kink}) into (\ref{e}) for $h=0$
we obtain for the energy per unit area of the $\pi$-Bloch wall
\begin{equation}
{\cal E}_K=\int_{-\infty}^{\infty} dz
\left( {d\phi_K \over d z} \right) ^2=2,
\label{ekink}
\end{equation}
where in the first step we have made use of the
fact that $\phi_K$ obeys
the ``energy conservation"
(\ref{econs}) with $C=1/2$ and $h=0$.

For $h\neq 0$, the degeneracy between the two anisotropy minima at
$(\theta,\phi)=(\pi/2,\pi)$ and $(\pi/2,0)$ is lifted.
Consequently, single Bloch walls cannot exist any more.
Instead two different types of Bloch wall pairs arise
which are discussed in the next two subsections.

\subsection{Untwisted domain wall pairs}

For $0<h<1$, the boundary conditions $\partial _z
\phi(\pm\infty)=0$,
$\phi(\pm\infty)=\pi$, imply that $C=1/2 -h$.
The integration  of (\ref{econs}) then yields the ``nucleus"
\cite{prlnucl,broz}
\begin{eqnarray}
\phi_s(z)&=&2 \arctan
\left( {\cosh z/\delta_s \over \sinh{R_s}}
\right),\nonumber\\ \theta_s&=&\pi/2.
\label{nucl}
\end{eqnarray}
As we shall see in Sec. V, the configuration
(\ref{nucl}) represents a {\it saddle point}
of the  energy since it is unstable for all values
$0<h<1$ of the external field.  Since it has exactly
one unstable mode, it represents a critical
nucleus for magnetization reversal.
The integration constant in (\ref{nucl}) is chosen such that the
symmetry center is located at $z=0$. Note, however, that
the continuous degeneracy of (\ref{nucl}) with respect to
translations will give rise to a zero energy (Goldstone)
mode in the fluctuation spectrum, quite analogous to the
case of the $\pi$-Bloch wall above. In (\ref{nucl})
we have introduced the ``radius" $R_s$ of the
untwisted domain wall pair. $R_s$ is related to the
 external field $h$ and the width $\delta_s$ as follows
\begin{equation}
h=\mathop{\rm sech}\nolimits^2 R_s,\qquad \delta_s=\coth R_s.
\label{nuclh}
\end{equation}
The nucleus may also be written as a superposition of
two  {\it untwisted} $\pi$-Bloch walls (\ref{kink}) centered at
$z/\delta_s=\pm R_s$,
\begin{equation}
\phi_s(z)=\phi_K\left(- {z\over\delta_s}+R_s\right)+
\phi_K\left( {z\over\delta_s}+R_s\right).
\label{nuclsup}
\end{equation}
Note that this relation is exact for all $0<R_s<\infty$.
Eqs. (\ref{nucl}), (\ref{nuclsup}) thus
describe a domain with  magnetization oriented parallel
to the external field  which is delimited by a pair of
untwisted $\pi$-Bloch walls (cf. Fig. \ref{nucleus}).
The existence of this structure is due to the balance of
exchange and Zeeman energy. The exchange energy tends to
attract the untwisted domain walls, whereas the Zeeman
energy pulls them apart since it favors the
intermediate domain.
As is illustrated  by
(\ref{nuclh}), (\ref{nuclsup}), and Fig. \ref{nucleus},
the domain wall separation
tends to infinity for $h\to 0$ whereas for
$h\to 1$ the two oppositely twisted
domain walls almost annihilate each other
and the nucleus degenerates to an
infinitesimal deviation
from the uniform ``down" state $(\phi,\theta)=(\pi,\pi/2)$.

Using the parametrization (\ref{nuclh}),
the energy per unit area (\ref{e}) of the nucleus
relative to the  ``down" state takes the simple form
\begin{eqnarray}
 {\cal E}_s&\equiv& {\cal E}[\phi_s,\theta_s]-
{\cal E}[\phi=\pi,\theta={\pi\over 2}]
= \int_{-\infty}^{\infty}
dz \left( {d\phi_s\over dz}\right)^2\nonumber\\
&=&4\tanh R_s - 4R_s\mathop{\rm sech}\nolimits^2 R_s.
\label{Es}
\end{eqnarray}
In the first step we have used the first integral
(\ref{econs}) and the integration is most
easily performed with (\ref{nuclsup}).
The first term on the r.h.s. in (\ref{Es})
describes the deformation energy of the nucleus compared
to the uniform state in the absence of an external field.
The second term is the Zeeman energy $-{\cal M}_s h$.
The magnetic moment per unit area relative
to the down state is thus given by
\begin{equation}
{\cal M}_s=4R_s.
\label{nuclmom}
\end{equation}
The deformation energy
vanishes for $R_s\to 0$ reflecting the fact that untwisted
pairs of domain walls are attractive.
For $R_s\to\infty$, the energy converges to that of two
independent $\pi$-Bloch walls.

{}From (\ref{Es}) and (\ref{nuclh}) we may immediately
derive the (formal) susceptibility
\begin{equation}
\chi_s\equiv {d {\cal M}_s\over d h}={2\over h\sqrt{1-h}}.
\end{equation}
This  susceptibility has only formal character,
since as we shall see below,
the nucleus is unstable for all values of the external
field $0<h<1$.

\subsection{Twisted domain wall pairs}

For $0<h<\infty$, and for the boundary conditions
$\partial _z\phi(\pm\infty)=0$, $\phi(\pm\infty)=0$,
we have $C=1/2+h$.
Eq. (\ref{econs}) may then be integrated to yield the
$2\pi$-Bloch wall \cite{bishop,thomasmag,diss}
\begin{eqnarray}
\phi_b(z)&=&2 \arctan \left( {\cosh R_b \over
\sinh{z/\delta_b}}\right),\nonumber\\
\theta_b&=&\pi/2.
\label{bloch}
\end{eqnarray}
The integration constant has been chosen such
that the symmetry center is located at $z=0$
but as in the case of the nucleus, the
translational  degeneracy will give rise to a zero energy
(Goldstone) mode in the fluctuation spectrum.
The ``radius" $R_b$ of the
twisted domain wall pair is related to the external field $h$
and the characteristic width $\delta_b$ as follows
\begin{equation}
h=\mathop{\rm csch}\nolimits^2 R_b,\qquad \delta_b=\tanh R_b.
\label{blochh}
\end{equation}

The $2\pi$-Bloch wall (\ref{bloch}) may also be
written as a superposition of
two  {\it twisted} $\pi$-Bloch walls (\ref{kink}) located at
$z/\delta_b=\pm R_b$,
\begin{equation}
\phi_b(z)=\phi_K\left( {z\over\delta_b}-R_b\right)+
\phi_K\left( {z\over\delta_b}+R_b\right).
\label{blochsup}
\end{equation}
This relation is valid for all values of $R_b$. Eqns
(\ref{bloch}), (\ref{blochsup}) describe a pair of
$\pi$-Bloch walls located at $z/\delta=\pm R_b$
with equal relative sense of twist,
enclosing a domain of reversed magnetization
(cf. Fig. \ref{2pblochwall}).
This  structure is stabilized by the
 balance of Zeeman and exchange
energy. The Zeeman energy tends to
enlarge the domains oriented
parallel to the external field,
whereas the exchange energy pulls the twisted
domain walls apart.  As illustrated by Fig. \ref{2pblochwall} b),
the $2\pi$-Bloch wall decays for $h\to 0$  into two individual
$\pi$-Bloch walls with increasing separation,
whereas for $h\to \infty$ (Fig. \ref{2pblochwall} a)),
the two $\pi$-Bloch walls are squeezed and the transition
region becomes infinitesimally small.

The energy per area of the $2\pi$-Bloch wall
is given by
\begin{eqnarray}
{\cal E}_b&\equiv&{\cal E}[\phi_b,\theta={\pi\over 2}]-
{\cal E}[\phi=0,\theta={\pi\over 2}] =
\int_{-\infty}^{\infty}
dz \left({d\phi_b\over dz}\right)^2\nonumber\\
&=&4\coth R_b + 4R_b\mathop{\rm csch}\nolimits^2R_b.
\label{Eb}
\end{eqnarray}
where (\ref{econs}) and   (\ref{blochsup}) have been used.
The first and second term
on the r.h.s. in (\ref{Eb})  describe the
deformation energy of the $2\pi$-Bloch wall
relative to the uniform ``up"-state $\phi=0$
in the absence of an external field,
and the Zeeman energy,  respectively.
The magnetic moment per unit area relative to the
up state is thus given by
\begin{equation}
{\cal M}_b=-4R_b.
\label{blochmom}
\end{equation}
Note that the deformation energy in (\ref{Eb})
diverges for $R_b\to 0$  (i.e., $h\to\infty$),
i.e.  a compression of the
$2\pi$-Bloch wall to zero width is connected with an
infinite increase in exchange energy. For $R_b \to\infty$
 (i.e. $h\to 0$) the deformation energy tends to
that of two single $\pi$-Bloch walls and the Zeeman
energy becomes zero. With (\ref{Eb}) and
(\ref{blochmom}) we obtain the susceptibility
\begin{equation}
\chi_b\equiv {d {\cal M}_b\over d h}
={2\over h\sqrt{1+h}}.
\end{equation}
For large external fields this susceptibility
has a only formal meaning, since the $2\pi$-Bloch wall
can become unstable for $h\raise1pt\hbox{$>$}\lower3pt
\hbox{\llap{$\sim$}} Q^{-1}/3$ as we shall see in Sec. V.

\section{fluctuations}

To investigate the stability of the structures
presented in the last section, we perform an
expansion around a given easy-plane configuration
$(\phi_0(z),\theta=\pi/2)$ as follows
\begin{eqnarray}
\phi(z)&=&\phi_0(z)+\varphi(z),\nonumber\\
\theta(z)&=&\pi/2-p(z),
\label{fluctu}
\end{eqnarray}
where $|\varphi|, |p| \ll 1$. First, we shall
review the fluctuations around the
$\pi$-Bloch wall  because of their close
relation to the fluctuations of the
$2 \pi$-Bloch wall and the nucleus.

Inserting (\ref{fluctu}) with $\phi_0=\phi_K$
into (\ref{e})  for $h=0$ we obtain up to
second order in $\varphi(z)$ and  $p(z)$:
\begin{equation}
{\cal E}^{(2)}={\cal E}^K +  {1\over 2} \int_{-\infty}^{\infty} \!\!\! dz
\;\varphi \,{\cal H}^K \varphi +
{1\over 2} \int_{-\infty}^{\infty} \!\! \! dz\; p\,
({\cal H}^K +Q^{-1})p,
\label{E2k}
\end{equation}
where ${\cal E}^K$ is the Bloch wall energy  (\ref{ekink}).
No first order term in the fluctuations  is present in (\ref{E2k})
since $\phi_K$  obeys the Euler-Lagrange equations
(\ref{EL}) with $h=0$. The operator ${\cal H}^K$ is defined as
\begin{equation}
{\cal H}^K = -{d^2 \over d z^2} + 1 -2
\mathop{\rm sech}\nolimits^2 z.
\label{Hk}
\end{equation}
The potential appearing in (\ref{Hk}) belongs to the family of
reflectionless potentials which
are of the form $-m(m+1)
\mathop{\rm sech}\nolimits^2 z$ ($m$
an integer), and  which are discussed in the appendix.
 The eigenvalue problem of  (\ref{Hk})  is
\begin{equation}
{\cal H}^K \chi^K_\nu (z) = E_\nu^K\chi^K_\nu(z),
\qquad \nu=0,k.
\label{Hkev}
\end{equation}
There is one bound state with zero energy
\begin{equation}
\chi^K_0 (z) ={1\over\sqrt{2}}
\mathop{\rm sech}\nolimits z,\quad
E_0^K=0,
\label{Hkb}
\end{equation}
and there are running (spin-wave) states
\begin{eqnarray}
\chi_k^K (z)&=&{1\over \sqrt{2\pi(1+k^2)}}\left[-ik +
\tanh z\right] e^{ikz},\nonumber\\
E_k^K&=&1+k^2.
\label{Hkk}
\end{eqnarray}
The easy-axis anisotropy leads to the gap 1
in the spin-wave spectrum (\ref{Hkb}), (\ref{Hkk})
while (\ref{E2k}) shows that the hard-axis anisotropy
gives rise to the ``mass" $Q^{-1}$ of the $p$-fluctuations.
Since $\chi_0^K$ is nodeless and thus represents
the ground state of ${\cal H}_K$, all eigenvalues
of ${\cal H}_K+Q^{-1}$ are positive.  Therefore all
fluctuations around a $\pi$-Bloch wall have positive
energy except for the zero energy mode
$(\varphi,p)=(\chi^K _0(z),0)$.
This mode corresponds to a rigid translation
of the Bloch wall:  Taking the derivative
of (\ref{econs}) for $h=0$ we obtain
${\cal H}^K d\phi_K/dz=0$ and therefore
$\chi^K _0\propto d\phi_K/dz$.
We conclude that in the absence of an external field the
static kink is stable with respect to planar distortions
except for rigid translations which involve zero energy.
This result was first obtained by Winter \cite{Winter}.
We now proceed with a discussion of fluctuations
of the nucleus and the $2\pi$-Bloch wall.

\subsection{Nucleus}

Inserting (\ref{fluctu}) with $\phi_0=\phi_s$
into (\ref{e}) and
evaluating ${\cal E}[\phi,\theta]-{\cal E}[\phi=\pi,\theta=
{\pi\over 2}]$ to $2^{nd}$ order in  $\varphi$ and $p$ we obtain
\begin{equation}
{\cal E}_s^{(2)}\equiv {\cal E}_s +{1\over 2}
\int_{-\infty}^{\infty} dz\;
\varphi
\,{\cal H}^{s\varphi} \varphi +{1\over 2} \int_{-\infty}^{\infty} dz\;
p\,{\cal H}^{sp} p,
\label{e2s}
\end{equation}
where ${\cal E}_s$ is given by (\ref{Es}). The first order
term in the fluctuations is absent since  $\phi_s$ satisfies the
Euler-Lagrange
equations  (\ref{EL}).
The operators ${\cal H}^{s\varphi}$ and
${\cal H}^{sp}$ are defined as
\begin{equation}
{\cal H}^{s\varphi}=-{d^2\over d z^2} +
2\cos^2\phi_s + \mathop{\rm sech}\nolimits^2R_s\cos\phi_s
-1,
\label{Hsf1}
\end{equation}
\begin{eqnarray}
{\cal H}^{sp}=-{d^2\over d z^2} &+&2\cos^2\phi_s  +
3\mathop{\rm sech}\nolimits^2R_s\cos\phi_s\nonumber\\
&+&2\mathop{\rm sech}\nolimits^2R_s-1+Q^{-1},
\label{Hsp1}
\end{eqnarray}
with
\begin{equation}
\cos\phi_s =  {\sinh^2  R_s-  \cosh^2  \left( {z/\delta_s}\right)
                \over\sinh^2 R_s   + \cosh^2  \left({z/\delta_s}\right)}.
\label{cosfs}
\end{equation}
This form  of the fluctuation operators is rather
involved.   Since the
nucleus can be represented as a superposition (\ref{nuclsup})
of untwisted $\pi$-Bloch walls, we expect these
operators to contain potentials  of the form (\ref{Hk})
for each of the constituents of the nucleus.  In fact
(\ref{Hsf1}) and (\ref{Hsp1})  allow for the much
simpler representation
\begin{equation}
{\cal H}^{s\varphi}=-{d^2\over d
z^2}+\delta_s^{-2}V_-\left({z\over\delta_s},R_s\right),
\label{Hsf}
\end{equation}
\begin{equation}
{\cal H}^{sp}=-{d^2\over d z^2}+\delta_s^{-2}V_
+\left({z\over\delta_s},
R_s\right)+
Q^{-1},
\label{Hsp}
\end{equation}
where the potentials $V_\pm$  are given by
\begin{eqnarray}
V_\pm (\zeta,R)=1&-&2\mathop{\rm sech}
\nolimits^2(\zeta+R)
-2\mathop{\rm sech}\nolimits^2(\zeta-R)\nonumber\\
&\pm& 2\mathop{\rm sech}\nolimits(\zeta+R)
\mathop{\rm sech}\nolimits(\zeta-R).
\label{V}
\end{eqnarray}
The second and third term on the r.h.s of (\ref{V}) are the
potentials  (\ref{Hk}) of two noninteracting $\pi$-Bloch walls
located at $z/\delta_b=\pm R_b$.  The  last term, which
vanishes for $R_s\to\infty$,
describes the interaction of the two $\pi$-Bloch walls
and is thus sensitive to their relative sense of twist.
The constant $Q^{-1}$ in (\ref{Hsp}) is due to the hard-axis
anisotropy and leads,  in analogy to the $\pi$-Bloch wall,
to a finite mass of  fluctuations out of the easy plane.
The corresponding eigenvalue problems are
\begin{eqnarray}
{\cal H}^{s\varphi}\chi^{s\varphi}_\nu (z,R_s) &=
&E^{s\varphi}_\nu(R_s)\chi^{s\varphi}_\nu(z,R_s),
\label{Hsfev}\\
{\cal H}^{sp}\chi^{sp}_\nu (z,R_s)
&=&E^{sp}_\nu(R_s)\chi^{sp}_\nu(z,R_s).
\label{Hpfev}
\end{eqnarray}
where the index $\nu$ denotes bound states and scattering states.
An analytical solution of these eigenvalue problems
seems only possible in the limiting cases $R_s\to 0$ and
$R_s\to\infty$.   However, one bound state of
${\cal H}^{s\varphi}$,
the zero energy state, can be derived immediately by
taking advantage of the continuous degeneracy of (\ref{nucleus})
with respect to translations. Taking the $z$-derivative of
(\ref{EL}) at $\phi=\phi_s$, we obtain with (\ref{Hsf1})
${\cal H}^{s\varphi} d\phi_s/ dz=0$, and therefore
\begin{eqnarray}
\chi^{s\varphi}_1&\propto&{d\phi_s\over dz}=\delta_s^{-1}
\left\{\mathop{\rm sech}\nolimits({z\over \delta_s}+R_s)
-\mathop{\rm sech}\nolimits({z\over\delta_s}-R_s) \right\},
\nonumber\\
E^{s\varphi}_1&=&0.
\label{nzero}
\end{eqnarray}
The antisymmetry of the zero mode $\chi^{s\varphi}_1$ with respect to
$z$ is a consequence of the opposite relative sense of
twist of the two $\pi$-Bloch walls in (\ref{nuclsup}).
The remaining bound state energies and
the scattering phase  shifts will be investigated
analytically and numerically in the next section.

\subsection {$2\pi$ Bloch wall}

Inserting (\ref{fluctu})  with $\phi_0=\phi_b$  into (\ref{e})
we obtain for ${\cal E}[\phi,\theta]-{\cal E}[\phi=0,
\theta={\pi\over 2}] $
to $2^{\rm nd}$ order in  $\varphi$ and $p$
\begin{equation}
{\cal E}_b^{(2)}\equiv{\cal E}_b +
{1\over 2} \int_{-\infty}^{\infty} dz\; \varphi \,{\cal H}^{b\varphi}
\varphi +{1\over 2}
\int_{-\infty}^{\infty} dz\; p\,{\cal H}^{bp} p,
\label{e2b}
\end{equation}
with ${\cal E}_b$ given by (\ref{Eb}).
The operators ${\cal H}^{b\varphi}$ and ${\cal H}^{bp}$
are defined as
\begin{equation}
{\cal H}^{b\varphi}=-{d^2\over d z^2} +2\cos^2\phi_b  +
\mathop{\rm csch}\nolimits^2 R_b\cos\phi_b -1,
\label{Hbf1}
\end{equation}
\begin{eqnarray}
{\cal H}^{bp}=-{d^2\over d z^2}&+&2\cos^2\phi_b  +
3  \mathop{\rm csch}\nolimits^2 R_b\cos\phi_b\nonumber\\
&-&2  \mathop{\rm  csch}\nolimits^2 R_b-1+Q^{-1},
\label{Hbp1}
 \end{eqnarray}
with (\ref{bloch})
\begin{equation}
\cos\phi_b =   {\sinh^2  \left(  {z/\delta_b}  \right) -
\cosh^2R_b\over\sinh^2\left({z/\delta_b}\right)+
\cosh^2R_b}.
\label{cosfb}
\end{equation}
The operator (\ref{Hbf1}) is identical to that describing the
fluctuations around a kink in the double sine-Gordon
model \cite{Campbell}. In analogy to  the nucleus,
(\ref{Hbf1}) and (\ref{Hbp1}) allow for a much simpler
representation
\begin{equation}
{\cal H}^{b\varphi}=-{d^2\over d z^2}+\delta_b^{-2}V_+
\left({z\over\delta_b},R_b\right),
\label{Hbf}
\end{equation}
\begin{equation}
{\cal H}^{bp}=-{d^2\over d z^2}+\delta_b^{-2}V_-
\left({z\over\delta_b},R_b\right)+Q^{-1},
\label{Hbp}
\end{equation}
where the potentials $V_\pm$  are given by (\ref{V}).
The representation (\ref{Hbf})
has also been obtained  by Sodano {\it et al.} \cite{Sodano}
in the discussion of kinks in the double sine-Gordon model.
It is instructive to compare (\ref{Hbf}) and (\ref{Hbp}) with
(\ref{Hk}):  The  $2^{nd}$ and
$3^{rd}$ term on the r.h.s. of (\ref{V}) are the
potentials of the noninteracting domain walls located at
$z/\delta_b=\pm R_b$. The last term describes the
interaction of the two domain walls and vanishes for
$R_b\to\infty$.  The constant $Q^{-1}$ in (\ref{Hbp})
is due to the hard-axis anisotropy  and leads
to a finite mass of  out of easy-plane fluctuations.

We write the eigenvalue problem of (\ref{Hbf}) and (\ref{Hbp})
in the following form:
\begin{eqnarray}
{\cal H}^{b\varphi}\chi^{b\varphi}_\nu (z,R_b)
&=&E^{b\varphi}_\nu(R_b)\chi^{b\varphi}_\nu(z,R_b),\\
{\cal H}^{bp}\chi^{bp}_\nu (z,R_b) &=&E^{b p}_\nu(R_b)
\chi^{bp}_\nu(z,R_b),
\end{eqnarray}
The index $\nu$ denotes bound states and scattering states.
Again, an analytic solution of these eigenvalue problems
 seems only possible
in the limiting cases $R_s\to 0$ and $R_s\to\infty$.
In analogy to the nucleus, one bound state of
${\cal H}^{s\varphi}$ can be derived immediately.
Taking the $z$-derivative of (\ref{EL}) at
$\phi=\phi_s$ we obtain ${\cal H}^{s\varphi}
 d\phi_s/dz=0$  and therefore
\begin{eqnarray}
\chi^{b\varphi}_0&\propto&{d\phi_b\over dz}=
\delta_b^{-1}\left\{\mathop{\rm sech}
\nolimits({z\over \delta_b}+R_b)
+\mathop{\rm sech}\nolimits({z\over\delta_b}-R_b)
\right\},\nonumber\\
E^{b\varphi}_0&=&0.
\label{bzero}
\end{eqnarray}
The symmetry of $\chi^{b\varphi}_0$ with respect to
$z$ reflects the equal sense
of twist of the two domain walls in (\ref{bloch}).

\section{Instabilities}

We are now in a position to state one of the central
results of this paper. Comparing (\ref{Hbf}), (\ref{Hbp})
with  (\ref{Hsp}) and (\ref{Hsf}) we infer the
remarkable  connection
\begin{eqnarray}
{\cal H}^{sp}(z, R)&=&\left({\delta_b\over\delta_s}\right)^2
{\cal H}^{b\varphi}({\delta_b\over\delta_s}z, R)+Q^{-1},
\label{Hplink}\\
{\cal H}^{bp}(z, R)&=&\left({\delta_s\over\delta_b}\right)^2
{\cal H}^{s\varphi}({\delta_s\over\delta_b}z, R)+Q^{-1}.
\label{Hbplink}
\end{eqnarray}
Here, for clarity,  the
notation ${\cal H}^{s\varphi}(z,R)\equiv -d^2/dz^2 +
\delta_s^{-2} V_-({z\over\delta_s},R)$, $\delta_s=\coth R$,
has been used and analogously for the remaining operators.
Eqns (\ref{Hplink}), (\ref{Hbplink}) show
that the fluctuations around the
$2\pi$-Bloch wall and around the nucleus are governed
up to rescaling by the same set of operators.
Consequently the  eigenvalues are related by
\begin{eqnarray}
E^{sp}_\nu(R)&=&\left({\delta_b\over\delta_s}\right)^2
E^{b\varphi}_{\nu'}(R)+Q^{-1},
\label{Eplink}\\
E^{b p}_{\nu'}(R)&=&\left({\delta_s\over\delta_b}
\right)^2E^{s\varphi}_\nu(R)+Q^{-1},
\label{Ebplink}
\end{eqnarray}
and the eigenfunctions obey
\begin{eqnarray}
\chi^{sp}_\nu(z,R)&=&\chi^{b\varphi}_{\nu'}
({\delta_b\over\delta_s}z,R),
\label{cplink}\\
\chi^{bp}_{\nu'}(z,R)&=&\chi^{s\varphi}_\nu
({\delta_s\over\delta_b}z,R),
\label{cbplink}
\end{eqnarray}
where for  bound states  $\nu={\nu'}$ and for
scattering states  $\nu=k$, ${\nu'}=(\delta_s/\delta_b) k$.
The continuum eigenvalues are  defined
as $E_k^{j\varphi}=\delta_j^{-2}+k^2$,
$E_k^{jp}=Q^{-1}+E_k^{j\varphi}$
for $j=s,b$.
In (\ref{Hplink})-(\ref{cbplink}) we have
used
\begin{equation}
\delta_b/\delta_s=\tanh^2 R.
\end{equation}
The relations  (\ref{Eplink}), (\ref{Ebplink})
together with (\ref{nzero}),  (\ref{bzero})   now allow us to
discuss instabilities of the nucleus and the $2\pi$-Bloch wall
in a simple and straightforward way.

The function $\chi^{b\varphi}_0$ as given in (\ref{bzero})
is symmetric and nodeless, and hence it represents the
ground state of ${\cal H}^{b\varphi}$ with  zero energy.
Except for this state, ${\cal H}^{b\varphi}$ has a strictly
positive spectrum and so has ${\cal H}^{sp}$, i.e.
\begin{equation}
E^{b\varphi}_\nu(R_b) \geq 0,\;\;E^{sp}_\mu(R_s) >0,
\label{Epos}
\end{equation}
for all $\nu$ and  $0<R_b,R_s<\infty$. It thus follows that i)
the $2\pi$-Bloch wall is stable with respect
to easy-plane fluctuations (neutrally stable with
respect to the zero mode), and ii) that the nucleus is
stable with respect to out of easy-plane fluctuations.

On the other hand, the function $\chi^{s\varphi}_1$
is antisymmetric  with one
node and thus represents the first excited state of
${\cal H}^{s\varphi}$. Since it has zero energy,
there  is exactly one nodeless, symmetric bound
state of {\it negative} energy, i.e.
\begin{equation}
E^{s\varphi}_0(R_s)<0,
\label{Eneg}
\end{equation}
for all $0<R_s<\infty$. The inequality (\ref{Eneg})
is the origin of the following instabilities:

{\it nucleus}:
Fluctuations in $\varphi$ direction exhibit
exactly one mode of negative energy
$E^{s\varphi}_0$. Since ${\cal H}^{sp}$ is positive
(cf. (\ref{Epos})), we conclude that {\it
there is exactly one unstable mode of the nucleus for
all values of $R_s$}.
Since $\phi_s$ is untwisted (i.e. $q(\phi_s)=0$), the
instability in $\varphi$ provides
an example of a topologically induced instability.

{\it $2\pi$-Bloch wall}:
The $2\pi$-Bloch wall is stable with respect to $\varphi$
fluctuations because  of (\ref{Epos}).
Since $q(\phi_b)=1$, this stability is of topological
origin.  However,  { \it an instability  against
out of easy-plane distortions occurs if}
\begin{equation}
E^{b p}_0(R_b)\equiv Q^{-1} -\coth^4 R_b|E^{s\varphi}_0(R_b)| <0.
\label{blochinst}
\end{equation}
where $R_b$ is related to the external field as
$h=\mathop{\rm csch}\nolimits^2 R_b$.
In (\ref{blochinst}) we have made use of
(\ref{Ebplink}).

Anticipating results of the next section for the
asymptotic behaviour of  the eigenvalues,
we obtain the following asymptotic behaviour
for this instability condition
\begin{eqnarray}
Q^{-1}&<&2 h,\quad h\ll 1,\label{blochinstsmall}\\
Q^{-1}&<&3h,\quad h\gg 1,\label{blochinstlarge}
\end{eqnarray}
The numerically evaluated
instability condition (\ref{blochinst}) is shown in Fig.
\ref{instab} together with its asymptotic behavior
 (\ref{blochinstsmall})  and  (\ref{blochinstlarge}) .
The instability of the $2\pi$ Bloch wall
is in accordance with the  result of Magyari and Thomas
\cite{thomasmag} who gave also an  improved analytical
estimate of the instability range for large $h$,
however they did not discuss the nucleus
and the relation of its fluctuations to
the $2\pi$-Bloch wall.

Since we have shown in this section that the
eigenfunctions and eigenvalues of  ${\cal H}^{bp}$ and
${\cal H}^{b\varphi}$ can be expressed by
those of ${\cal H}^{sp}$ and
${\cal H}^{s\varphi}$,  we may restrict ourselves
to a discussion of the latter operators in the following.

\section{Discussion of ${\cal H}^{\lowercase{{\rm s}\varphi}}$ and
${\cal H}^{\lowercase{{\rm s} p}}$ }

In this section we evaluate the eigenfunctions of
${\cal H}^{s\varphi}$, ${\cal H}^{sp}$ numerically
and provide  analytical results
in  the limits of large and small $R_s$.
We first discuss bound state energies
which are related to the stability properties
of the $2\pi$-Bloch wall and the nucleus.
In view of statistical mechanical approximations,
the scattering phase shifts of the  continuum
eigenfunctions are discussed.
Furthermore it is shown that the
appearance of zero energy resonances in
the spectrum require a subtle analysis of
the applicability  of analytical approximations.

\subsection{Bound states}

In the limit of large and small $R_s$,  the eigenvalue
problems of  ${\cal H}^{s\varphi}$ and ${\cal H}^{s p}$
can be solved exactly:

For {\it large} $R_s$, the potentials
$V_\pm({z\over \delta_s},R_s)$
decay into two {\it independent} wells of the form
$-2\delta_s^{-2}{\mathop{\rm sech}\nolimits}^2
({z\over \delta_s}\pm R_s)$
and we denote the corresponding operators by
$\hat{\cal H}^s$.   This limit of large $R_s$ is
sometimes also referred to as  ``thin-wall limit"\cite{coleman}.
The bound states of ${\cal H}^{s\varphi}$ and
${\cal H}^{s p}$ are then  given by the symmetric
and antisymmetric combinations of the bound
states of the single wells. For $R_s\to\infty$
we thus have $\chi_0^{s\varphi}\to\hat
\chi_0^{s\varphi}$ and  $\chi_1^{sp}\to\hat \chi_1^{sp}$,
where
\begin{eqnarray}
\hat\chi^{s\varphi}_0(z)&\propto&
{\mathop{\rm sech}\nolimits}({z\over \delta_s}+R_s)
+{\mathop{\rm sech}\nolimits}({z\over \delta_s}-R_s),
\label{cf0}\\
\hat\chi^{sp}_1(z)&\propto&
{\mathop{\rm sech}\nolimits}({z\over \delta_s}+R_s)-
{\mathop{\rm sech}\nolimits}({z\over \delta_s}-R_s).
\label{cp1}
\end{eqnarray}
Note that the r.h.s. of (\ref{cf0}) and (\ref{cp1})
are the exact zero energy eigenfunctions of
${\cal H}^{s p}-Q^{-1}$ and
${\cal H}^{s\varphi}$,  respectively.
Since for large $R_s$ these operators  differ by
a term ${\cal O}(e^{-2 R_s})$,  we obtain within first order
perturbation theory:
\begin{eqnarray}
&\hat E^{s\varphi}_0&(R_s)\simeq{(\hat\chi^{s\varphi}_0,
{\cal H}^{s\varphi}\hat\chi^{s\varphi}_0)\over (\hat\chi^{s\varphi}_0,
\hat\chi^{s\varphi}_0)}=\nonumber\\
&-&\delta_s^{-2}\left[{3\over\cosh^2 R_s}+{1\over\sinh^2 R_s}\;
{2 R_s-\sinh 2 R_s\over 2 R_s+\sinh 2 R_s}\right],
\label{Ef0pert0}\\
&\simeq&-8  e^{-2R_s}
\label{Ef0pert}
\end{eqnarray}
and
\begin{eqnarray}
&\hat E^{sp}_1&(R_s)\simeq
{(\hat\chi^{sp}_1,{\cal H}^{s p}\hat\chi^{sp}_1)\over (\hat\chi^{sp}_1,
\hat\chi^{sp}_1)}=\nonumber\\
&\delta_s^{-2}&\left[{3\over\sinh^2 R_s} + {1\over\cosh^2 R_s} \;
{2 R_s+\sinh2R_s\over 2 R_s-\sinh2 R_s}\right]+ Q^{-1},\label{Ep1pert0}\\
&\simeq&8  e^{-2R_s} + Q^{-1}
\label{Ep1pert}
\end{eqnarray}
where $(u,v)$ denotes the standard scalar product $\int dz u^{\ast} v$.

For {\it small} $R_s$,  we have ${\cal H}^{s\varphi}\to \bar {\cal H}
^{s\varphi}$
and  ${\cal H}^{sp}\to \bar {\cal H}^{sp}$ with
\begin{eqnarray}
\bar{\cal H}^{s\varphi}&=&-{d^2\over dz^2} +
\delta_s^{-2}[1-6{\mathop{\rm sech}\nolimits}^2
\left({z\over \delta_s}\right)],
\label{Hfbar}\\
\bar{\cal H}^{s p}&=&-{d^2\over dz^2} +
\delta_s^{-2}[1-2{\mathop{\rm sech}\nolimits}^2
\left({z\over \delta_s}\right)]+ Q^{-1}.
\label{Hpbar}
\end{eqnarray}
Both potentials (\ref{Hfbar}), (\ref{Hpbar}) belong
to the class of  reflectionless potentials
which are discussed in the appendix.

The (unnormalized) bound states of
$\bar{\cal H}^{s\varphi}$ and their energies  are given by
\begin{eqnarray}
\bar \chi^{s\varphi}_0(z)&=&{\mathop{\rm sech}\nolimits}^2{z\over \delta_s},
\quad \hfill
\bar E^{s\varphi}_0=-3 \delta_s^{-2},
\label{barEf0}\\
\bar \chi^{s\varphi}_1(z)&=&
{\mathop{\rm sech}\nolimits}{z\over \delta_s}\tanh{z\over \delta_s},
\quad\hfill\bar E^{s\varphi}_1=0,
\label{barEf1}
\end{eqnarray}
and the spin wave states read
\begin{eqnarray}
\bar \chi^{s\varphi}_k(z)&=&\left( 3\tanh^2{z\over \delta_s} -
3 ik \delta_s\tanh{z\over \delta_s} -1 -
(k \delta_s)^2 \right) e^{ikz},
\nonumber\\
\bar E^{s\varphi}_k&=& \delta_s^{-2}+k^2.
\label{barEfk}
\end{eqnarray}

The operator $\bar{\cal H}^{s p}$ is up to rescaling
analogous to the operator (\ref{Hk}) which describes the
fluctuations around a single $\pi$-domain wall.
It has one bound state
$\bar\chi^{sp}_0(z)={\mathop{\rm sech}\nolimits}(z/\delta_s)$
with energy $\bar E^{sp}_0= Q^{-1}$, and spin wave states,
\begin{eqnarray}
\bar\chi^{sp}_k(z)&=&\left(-ik \delta_s+\tanh{z\over \delta_s}\right)
e^{ikz},\nonumber\\
\bar E^{sp}_k&=& Q^{-1}+\delta_s^{-2}+ k^2.
\label{barEpk}
\end{eqnarray}
In  Eqs (\ref{Hfbar})-(\ref{barEpk}) we have to put $\delta_s=R_s$
in order to be consistent with the   terms neglected in the
derivation of  $\bar{\cal H}^{sp}$ and $\bar{\cal H}^{s\varphi}$.

We are now in a position to verify the asymptotic behavior
of the  instability threshold of the $2\pi$-Bloch-wall.
Inserting (\ref{Ef0pert}), (\ref{barEf0}) into (\ref{blochinst})
we obtain (\ref{blochinstsmall}), (\ref{blochinstlarge}), respectively.

For arbitrary $R_s$, the bound state energies of
${\cal H}^{s\varphi}$ and ${\cal H}^{s p}$ have
been evaluated numerically and the results are
summarized in Fig. \ref{boundstates}.
The values of the asymptotic formulas (\ref{Ef0pert0}) and
(\ref{Ep1pert0}) are represented by dashed lines.
Note that they are accurate  for values as small as
$R_s\simeq 1.5$.  The operator
${\cal H}^{s\varphi}$ has  three bound states,
the ground state  of negative  energy $E^{s\varphi}_0$,
the zero-mode with $E_1^{s\varphi}=0$ and
a weakly bound state whose energy is always within 1\% of the
continuum threshold  according to numerical calculations.
For  applications such as the evaluation of nucleation rates we can therefore
use
\begin{equation}
E_2^{s\varphi}\simeq \delta_s^{-2}.
\end{equation}
This bound state  does not seem to be a numerical artifact
since its existence also follows from the  long-wavelength
behaviour of the scattering phase shifts $\Delta^{s\varphi}_{(e)}$
as we shall see in the next section.
The ground state wave function $\chi_0^{s\varphi}$ can  be considered
as an internal ``breathing" mode of the nucleus,
corresponding to an expansion or shrinking, depending on the sign of
$\chi_0^{s\varphi}$. Note, however,
that according to (\ref{cf0}), a strict equality $\chi_0^{s\varphi}
\propto d\phi_s/dR_s$ only holds in the limit $R_s\to\infty$.
The operator ${\cal H}^{s p}$ always has two bound states.
The  ground state with  constant  energy
$E_0^{sp}=Q^{-1}>0$ has its origin in the Goldstone mode of the
$2\pi$-Bloch wall while the excited state $\chi_1^{sp}$ of energy
$E_1^{sp}$ is related to  the ``breathing" mode of
the $2\pi$-Bloch wall.

Comparing the previous analytical discussion with these
numerical  results we are left with a paradox.
$\bar{\cal H}^{s\varphi}$  and  $\hat{\cal H}^s$ both have
{\it two} bound states, whereas numerical calculations reveal
the existence of {\it three}  bound states of ${\cal H}^{s\varphi}$.
Similarly, $\bar{\cal H}^{sp}$ has {\it one} bound state whereas
$\hat{\cal H}^s$ and ${\cal H}^{sp}$ have {\it two}  bound states.
The resolution of this paradox lies in the fact that
each of the operators obtained in the limits
$R_s\to 0, \infty$ exhibits a zero energy bound state.
Any increase in the potential strength thus leads to an
additional bound state which is precisely the reason for the
excess bound states of ${\cal H}^{s\varphi}$ and ${\cal H}^{sp}$.
The two well approximation  $\hat{\cal H}^s$  has  the same
number of bound states as ${\cal H}^{sp}$
but an additional zero energy resonance.
As $R_s$ becomes finite, the zero energy
resonance of  $\hat{\cal H}^s$ is shifted  into the
continuum.

\subsection {Scattering Phase Shifts}

The knowledge of scattering phase shifts is of importance
for statistical mechanical applications. In particular, the
results  of the present section will be used in a forthcoming
article \cite{braunnucl} on nucleation of domain wall pairs.
The scattering phase shifts
$\Delta_{(o)}^{si}$ ($\Delta_{(e)}^{si}$)
of the odd (even) eigenfunctions
$\chi_{k,(o)}^{si}$, ($\chi_{k,(e)}^{si}$)
of the operators
${\cal H}^{si}$, $i=\varphi,p$ are defined as follows:
\begin{eqnarray}
\chi_{k,(e)}^{si}(z\to\pm\infty,R_s)
&\propto&\cos\left[
kz\pm \Delta_{(e)}^{si}
(k,R_s)/ 2\right],
\label{evenphase}\\
\chi_{k,(o)}^{si}(z\to\pm\infty,R_s)&\propto&\sin\left[ kz\pm
\Delta_{(o)}^{si} (k,R_s)/ 2\right],
\label{oddphase}
\end{eqnarray}
where $ i=\varphi,p$.  It is sufficient to restrict our
considerations to the phase shifts of ${\cal H}^{s\varphi}$, since
according to (\ref{cplink}), (\ref{cbplink}) we have
\begin{eqnarray}
\Delta^{b\varphi}_{(j)}(k,R)&=&\Delta^{sp}_{(j)}({\delta_b\over\delta_s}k,R),\\
\Delta^{bp}_{(j)}(k,R)&=&\Delta^{s\varphi}_{(j)}({\delta_b\over\delta_s}k,R),
\end{eqnarray}
where $j=e, o$.

For {\it large} $R_s$, the potentials
$-2\delta_s^{-2}{\mathop{\rm sech}\nolimits}^2
(z/ \delta_s\pm R_s)$ act as independent scattering
centers, each contributing a phase shift
$2 \arctan(1/k\delta_s)$. Therefore we have
\begin{equation}
\hat\Delta^s(k)=4\arctan{1\over k \delta_s}.
\label{Dhat}
\end{equation}
For {\it small} $R_s$, the continuum eigenfunctions
(\ref{barEfk}) and
(\ref{barEpk}) of $\bar{\cal H}^{s\varphi}$ and
$\bar{\cal H}^{s p}$ lead to \cite{continuity}
\begin{eqnarray}
\bar \Delta^{s\varphi} (k)
&=& 2 \arctan  {3 k \delta_s \over
(k \delta_s)^2 - 2 },\label{Dfbar}\\
\bar\Delta^{sp} (k)
&=& 2\arctan {1\over k \delta_s}.\label{Dpbar}
\end{eqnarray}
Eqns.  (\ref{Dhat})-(\ref{Dpbar}) do not distinguish  between
odd and even parity eigenfunctions.

It is a surprising fact that some of the scattering
phases $\Delta_{(j)}^{si}$, $i=\varphi,p$, $\; j=e,o$
{\it do not converge uniformly}
to  (\ref{Dhat}-\ref{Dpbar})  in the limits
$R_s\to 0$ and $R_s\to \infty$,
respectively. Numerical calculations show that
(cf. Figs. \ref{Hfphases},\ref{Hpphases}),
\begin{eqnarray}
\Delta^{s\varphi}_{(e)}(k\to 0,R_s)&=&3\pi ,\nonumber\\
\Delta^{sp}_{(e)}(k\to 0,R_s)&=&\pi, \nonumber\\
\Delta^{sp}_{(o)}(k\to 0,R_s)&=&2\pi,
\label{dk0}
\end{eqnarray}
for all $R_s$.
Eqns (\ref{Dhat})-(\ref{Dpbar}), however, deliver  the relations
\begin{eqnarray}
\bar \Delta^{s\varphi}(k\to 0)&=&2\pi,\nonumber\\
\hat\Delta^{s}(k\to 0)&=&2\pi,\nonumber\\
\bar\Delta^{sp}(k\to 0)&=&\pi.
\label{dk0R0}
\end{eqnarray}
This discrepancy has the same roots as the paradox encountered
in the previous subsection, namely the existence of zero energy
resonances. This is elucidated by the widely unknown
1D version of Levinson's theorem \cite{Barton} which relates
the long wavelength limit of the phase shifts with
the number of bound states:

The {\it odd-parity} wavefunctions behave like in a
3D scattering problem:
\begin{equation}
\Delta^{si}_{(o)}(k\to 0) = 2\pi N^i_{(o)}, \quad i=\varphi,p,
\label{doddk0}
\end{equation}
where $N^i_{(o)}$ is the number of odd-parity bound states of
${\cal H}^{si}$. In the absence of zero energy resonances,
the scattering phase shifts of {\it even-parity} wavefunctions obey
a different relation
\begin{equation}
\Delta^{si}_{(e)}(k\to 0) = 2\pi (N^i_{(e)}-{1\over 2}),\quad
i=\varphi,p
\label{devenk0}
\end{equation}
where $N^i_{(e)}$ is the number of even-parity bound states of
${\cal H}^{si}$.
According to the previous subsection, we have for all values of $R_s$
\begin{eqnarray}
N_{(e)}^p&=&N_{(o)}^p=N_{(o)}^\varphi=1,\\
N_{(e)}^\varphi&=&2.
\end{eqnarray}
This shows that the $k\to 0$ behavior of the scattering phase shifts
 (\ref{dk0}) is in complete agreement
with the number of bound states of ${\cal H}^{s\varphi}$ and
${\cal H}^{s p}$ as evaluated in
the previous subsection.

If zero energy resonances are present and if the potential
is {\it reflectionless}, Eqs. (\ref{doddk0}), (\ref{devenk0})
have to be replaced
\cite{Barton} by the {\it parity independent} expression
\begin{equation}
\Delta^{si}_{(j)}(k\to 0) = \pi\left ( N^i_{(e)}+ N^i_{(o)}\right),
\label{scatfree}
\end{equation}
where $i=\varphi,p$ and $j=e,o$. Eq. (\ref{scatfree})
relates  (\ref{dk0R0}) to the number of
bound states of the reflectionless operators $\bar{\cal H}^{s\varphi}$,
$\bar{\cal H}^{s p}$, and
$\hat {\cal H}^s$ as given in (\ref{Ef0pert})-(\ref{barEpk}).
Levinson's theorem thus relates the nonuniform convergence of the
scattering phase shifts towards  $\bar\Delta^{s\varphi}$,
$\bar\Delta^{sp}$, $\hat\Delta^s$ to the appearance
of zero energy resonances in
$\bar{\cal H}^{s\varphi}$, $\bar{\cal H}^{s p}$, and $\hat{\cal H}^s$.
How this subtlety affects statistical
mechanical considerations, will be discussed in a
forthcoming paper \cite{braunnucl}.

The short wavelength behaviour of the scattering phase shifts can
be described  within Born's approximation \cite{Joachain}.
For an operator $-d^2/dz^2+V(z)$ with a symmetric potential
($V(z\to\pm\infty)=0$), the phase shift is given by:
\begin{equation}
\tan{\Delta^{si}_{(j)}(k)\over 2}\simeq
-{1\over 2 k}\int_{-\infty}^\infty dz \,V(z)
\sin^2(kz).
\label{born1}
\end{equation}
If $k^{-1}$ is much smaller than variations in $V(z)$, we can use
$\sin^2(kz)\simeq1/2$ and after insertion of
$V_\pm(z/\delta_s,R_s)-\delta_s^{-2}$ (cf. (\ref{V}))
into (\ref{born1})  we obtain
\begin{equation}
{\Delta^{si}_{(j)}(k)\over 2}\simeq {1\over k \delta_s} [2\mp{R_s\over
\sinh R_s\cosh R_s}], \quad {\rm
for}\;\; k \delta_s\gg 1,
\label{born3}
\end{equation}
where the upper sign refers to the $i=p$ and the $\tan$-function
has been replaced by its argument.
Finally it is interesting to note that for $\bar{\cal H}^{s p}$ the
Born approximation  (\ref{born1}) with $\sin^2(kz) \to 1/2$
coincides with the exact result (\ref{Dpbar}).

\section{$2\pi$ Bloch  Walls in  Thin Films}

The results of the previous sections are
rigorous within the 1D model of a biaxial ferromagnet which
contains exchange, local anisotropies and
the coupling to an external field.
While we have seen in Sec. II that
local demagnetizing effects  can be incorporated into the
model by a redefinition of the anisotropy constants,
one  might question the applicability
to thin films where nonlocal demagnetizing effects
are not a priori negligible.
Since the nonlocal demagnetizing interaction decays algebraically
while the exchange interaction between domain
walls decreases exponentially,
we expect the exchange interaction to be dominant and
thus our model to be adequate for small domain wall separations.

Indeed, it is the purpose of this section to show that for
sufficiently thin films and at external fields
below the threshold (\ref{blochinst}), twisted  domain wall pairs
may be brought sufficiently close such that the
exchange interaction between the individual domain walls
becomes important and nonlocal demagnetizing effects become
irrelevant. In this case our model adequately describes
the equilibrium separation of the walls.  We shall use c.g.s units
throughout this section.

To be specific, we choose coordinate axes as in
Figs. \ref{thinfilms} a), b)
and consider a film of thickness $D$ in $x$-direction
which extends infinitely
in the $y$-direction and has length $L$ in $z$-direction.
Further we assume the magnetization to be strictly
one-dimensional, i.e. ${\bf M}={\bf M}(z)$.
The  demagnetizing energy per area
${\cal E}_m=-(1/2D)\int dx\; dz\;
{\bf H}_m(x,z)\cdot {\bf M}(z)$
can then be cast into a very convenient form
due to Dietze and Thomas \cite{dietze}:
\begin{eqnarray}
{\cal E}_m &=& \int_{-L/2}^{L/2}\!\!dz
\;2\pi M_z^2(z) + \nonumber \\
&&{1\over D} \int_{-L/2}^{L/2}\!\!
dz \int_{-L/2}^{L/2} \!\!dz'\;
[M_x(z) M_x(z')-M_z(z)M_z(z')]\times\nonumber\\
 &&\ln\left(1 + {D^2\over (z-z')^2}\right).
\label{dt}
\end{eqnarray}
Eq.  (\ref{dt}) reduces to a simple hard-axis anisotropy
energy in the following two limiting cases:

For a film thickness $D$ smaller than the characteristic
length scale of ${\bf M}$, i.e. smaller than the domain wall width,
the integration over the relative coordinate in the
second term on the r.h.s. in (\ref{dt}) can be performed,
and Eq. (\ref{dt}) reduces to
\begin{equation}
{\cal E}_m=2\pi \int_{-L/2}^{L/2} dz \;  M_x^2(z).
\label{dmhardaxis}
\end{equation}
This has the form of a hard-axis anisotropy energy normal to the
film plane.

In the opposite limit of large $D\gg L$,
the second term on the r.h.s. of (\ref{dt}) tends to zero \cite{demag}
and we recover the result (\ref{Hm1d}). This form of the
demagnetizing energy is used for the description
of domain wall dynamics in moderately thin ($\simeq 1\mu m$)
rare earth garnet films \cite{Enzetal,sloncz}.

As the magnetostatic interaction
has the form of a hard-axis anisotropy in these limits,
the energy density (\ref{e}) can be used to describe
three distinct experimental configurations.
In addition to the bulk situation considered so far
(cf. Fig. \ref{thinfilms} a), it
describes configurations in thin films ($D/\sqrt{A/K_e}\leq 1$)
which are perpendicularly (Fig \ref{thinfilms}  b)
or in-plane (Fig \ref{thinfilms} c) magnetized
provided that the coordinate axes are chosen
appropriately. The results of the previous
sections thus hold for all configurations shown in  Fig.
\ref{thinfilms} in the limit of infinitesimally thin films.
In the following it is discussed how {\it nonlocal
demagnetizing fields}, i.e. the nonlocal contribution in
(\ref{dt}),  will affect these results for a film
of finite thickness.

\subsection{Perpendicularly Magnetized Films}

Consider a situation as in Fig. \ref{thinfilms} b)
which requires  the crystalline easy-axis anisotropy
to be larger than the demagnetizing energy, i.e.
$K_{e,{\rm cryst}}>2\pi M_0^2$, and the easy-axis to be oriented
perpendicularly to the film, a situation typically realized in bubble films or
in Barium-Ferrite.  To estimate the nonlocal demagnetizing
interaction for domain wall separations large compared to a domain
wall width, we consider the configuration
 \begin{equation}
{\bf M}^{(0)}({\bf r})=\left\{   \begin{array}{ll}
                                      \phantom{-}M_0{\bf e}_x   &
                                                  \mbox{for $L> |z|>a$}\\
                                       -M_0{\bf e}_x   &
                                            \mbox{for $ |z|\leq a$},
                                \end{array}
                              \right.
\label{filmdistr}
\end{equation}
for $0\leq x \leq D$, $-\infty < y < \infty$ and vanishing elsewhere.
The nonlocal magnetostatic interaction is due to the nonuniform
surface charge distribution  caused by  the reversed domain at
$|z|\leq a$ (cf. Fig \ref{thinfilms} b).
Inserting (\ref{filmdistr}) into (\ref{dt}) and performing
the limit $L\to\infty$ {\it after} evaluation of the integrals,
we obtain
\cite{westervelt}
\begin{eqnarray}
&{\cal E}&_m^{\rm perp}={\cal E}_m[{\bf M}^{(0)}]-
{\cal E}_m[M_0 {\bf e}_x]=
\nonumber\\
&-&4M_0^2 D \alpha \left\{ 4\tan^{-1}{1\over\alpha} +
2 \alpha \ln\alpha
+({1\over\alpha}-\alpha) \ln(1+\alpha^2) \right\},
\label{perpdem}
\end{eqnarray}
where $\alpha=2a/D$ is the width of the reversed
domain with respect to the film thickness.
Note that the r.h.s of (\ref{perpdem}) decreases with
increasing $\alpha$ and thus
favors an expansion of the reversed domain independent of
the relative twist of the domain walls. For the twisted
domain wall pair it thus competes with the repulsive
exchange interaction.

An external magnetic field $H_{\rm ext}$ along the positive
$x$-direction (cf. Fig. 8 b) will counteract this
magnetostatic repulsion. A relation for the
corresponding equilibrium width
is obtained by minimizing the energy (\ref{perpdem}) plus
the Zeeman energy of the intermediate domain,
$2 M_0 H_{\rm ext} D\alpha$,
with respect to $\alpha$ with the result
\begin{equation}
{H_{\rm ext}\over 4 \pi M_0}={2\over \pi} \tan^{-1} {1\over \alpha}
-{\alpha\over \pi} \ln \left(1+{1\over \alpha^2}\right).
\label{widthperp}
\end{equation}
Note that the relation (\ref{widthperp}) does not depend on
the absolute size of the reversed domain but only
on its relative size with respect to the film thickness.
Since on the other hand the mutual exchange repulsion
of twisted domain walls  depends on their absolute distance,
it can only  be observed if the equilibrium width
in Eq. (\ref{widthperp}) is small.  However, since
the external field must not exceed the instability threshold,
this can only be achieved in sufficiently thin films.

To investigate this effect quantitatively,
we have to compare the equilibrium width
$2 a=\alpha D$ of (\ref{widthperp}) with
the separation $2 a=2 R_b \delta_b \delta_0$
of the domain walls forming a $2\pi$- Bloch wall
(\ref{blochh}), (\ref{blochsup}) where $\delta_0=\sqrt{A/K_e}$
is the  static Bloch-wall width. It is
sufficient to look at small  $H_{\rm ext} M_0/(2 K_e)$
where this relation between equilibrium width and external
field can be expressed as
\begin{equation}
{H_{\rm ext} \over 4 \pi M_0}= {2 K_e\over  \pi M_0^2}
 \exp\left\{- {2 a\over \delta_0} \right\}.
\label{wallsep}
\end{equation}
This topological (exchange) interaction between
the domain walls thus decreases exponentially, while the
magnetostatic interaction (\ref{widthperp}) decays
algebraically, i.e. $H_{\rm ext}/4\pi M_0= D/(2\pi a)$,
for large $2a/D$. Since the r.h.s of (\ref{wallsep})
is proportional to the ``quality factor"
$K_e /( 2 \pi M_0^2)$ of the domain wall, the
exchange repulsion will manifest itself at
larger domain wall separations with
increasing quality factors and decreasing film thickness.

This is illustrated in Fig. \ref{balance} where
the solid lines show the exchange dominated
wall separation (\ref{wallsep}) whereas the dashed lines
represent (\ref{widthperp}). The experimentally
required field  to reach a certain wall separation
will follow the curve that has
the maximal value of $H_{\rm ext}$ consistent
with the sample parameters. Note that the
exchange repulsion between the domain walls has a
drastic effect. E.g.
for a  film with $D/\delta_0=0.1$ and $K_e/(2\pi M_0^2)=10$,
the external fields to establish a distance
of $7\delta$  predicted by (\ref{wallsep})
exceeds that of (\ref{widthperp}) by an
order of magnitude, or vice versa, the
wall separations differ by a factor of
3 for $H_{\rm ext}/(4\pi M_0)=0.01$.
This large discrepancy should be accessible to experimental
verification.

So far we did not discuss the case of untwisted domain
wall pairs. Despite the fact that they have been
shown to be unstable in Sec. V, Eq. (\ref{widthperp})
shows that such domain wall pairs can exist
in thin films due to the balance of demagnetizing  and
Zeeman energy provided they are well separated.
However,  the ansatz (\ref{filmdistr}) overestimates
demagnetizing  effects.  As the  wall separation
decreases, the untwisted walls annihilate each other
(cf. Fig. \ref{nucleus} a))
and the magnetostatic surface charges are drastically
reduced compared to those of (\ref{filmdistr}).
This  implies that the experimentally observed
separation of untwisted
wall pairs will not follow the dashed curve in Fig.
\ref{balance} down to vanishing $a$ but exhibit an instability
at  finite $a$. This instability
will occur at fields that are much smaller than
the instability threshold of the twisted domain wall pair.
We do not consider this case further
but conclude with the remark that a quantitative theory
can be obtained by improving  the ansatz (\ref{filmdistr})
by replacing it by the nucleus solution $(\phi_s,\theta_s)$.

\subsection{In-Plane Magnetized Films}

Consider the configuration shown in  Fig. \ref{thinfilms} c) with
a crystalline easy-axis in the film plane. All previous formulae
hold also for this configuration provided that we redefine
all coordinates appropriately, i.e. $(x,y,z) \to (z,x,y)$.
Since the magnetization configuration is exclusively in the
film plane, there are no induced magnetostatic surface charges
in a film that extends infinitely in $x$-direction
and there is no equilibrium domain formation in
the infinite film geometry.  However, we assume that domain
walls exist that have been created  e.g. by nucleation
at a sample end and/or nonuniform external fields.
If domain walls are present, a magnetostatic interaction arises
between the magnetostatic volume charges of the domain walls.
For sufficiently well separated  twisted domain wall pairs,
Eq. (\ref{blochsup})  takes the form
$M_y/M_0={\rm sech}((y+a)/ \delta_0) - {\rm sech}((y-a)/\delta_0)$.
Inserting this into (\ref{dt}) (with redefined
coordinate axes) we obtain for $a$ much larger than
a domain wall width $\delta_0$,
\begin{equation}
{\cal E}_m^{\rm ip}=2 D (\pi M_0)^2
\left({\delta_0 \over 2 a}\right) ^2.
\label{inplanedemag}
\end{equation}
 Eq. (\ref{inplanedemag}) is simply the
interaction energy of two infinitely long strings of dipoles
along $y$ with dipole moment per unit
length  $\mu_y=D \int dy M_y(y)=$ $D \delta_0 \pi M_0$.
The magnetostatic interaction between the walls is thus
repulsive for the twisted domain wall pair and competes with
the exchange interaction between the individual walls.
(For untwisted domain wall pairs the magnetostatic
as well as the exchange interaction would be attractive.)
An external magnetic field in $x$-direction  will
counteract this repulsion. The equilibrium distance
$2a$  between two domain walls
is  obtained  by minimization of  demagnetizing plus Zeeman energy,
${\cal E}_m^{\rm ip}$ $+  4 M_0 H_{\rm ext}a$,
with the result
\begin{equation}
{ H_{\rm ext} \over  4 \pi M_0   }={ \pi\over 2} {D\over \delta_0}
  \left( \delta_0 \over  2 a   \right)^3.
\label{widthinplane}
\end{equation}
In Fig. \ref{balance} b), this is compared with the wall
separation (\ref{wallsep}) which is predicted by our model.
Demagnetizing effects are obviously weaker than in
a perpendicularly magnetized film.  For films of thickness
$D=0.1\delta_0$  and  $K_e/(2\pi M_0^2)=10$,  the exchange
interaction dominates  the demagnetizing interaction already
at a domain wall
separation of $2a=13\delta_0$ which is surprisingly large
considering the exponential decrease of the exchange
interaction (\ref{wallsep}).   Note also that the external fields
which are required to achieve a domain wall distance of
 $6\delta_0$ differ by a factor of $100$.
Finally, we note that untwisted wall pairs in in-plane
magnetized films are never stable for the anisotropy configuration
shown in Fig. \ref{thinfilms} c).

\section{Conclusion}

In this paper we have discussed the stability of twisted
and untwisted domain wall pairs within a 1D model of
a ferromagnet. The fluctuations around these structures have been
shown to be described by the same operators.
By means of exactly known eigenfunctions
which are related to the Goldstone modes of these
structures it has been shown that untwisted domain wall pairs exhibit
exactly one unstable mode while twisted domain wall pairs are subject to
an instability at large external fields.

Furthermore we have argued that this model adequately
describes the separation of twisted domain walls
in ultrathin films and thus the above instability
should be observable.

Although untwisted domain walls are unstable within
the biaxial ferromagnet, they can exist in thin
films at large separations due to the long range magnetostatic
interaction. However, as a consequence of their
topological instability, the corresponding collapse
field will be much smaller than that of twisted domain
wall pairs. There are experimental \cite{niedoba}
and numerical \cite{humphrey} hints for this behavior,
but more systematic studies are required to
allow for a quantitative comparison with the present theory.

Another important aspect is the role of untwisted
domain wall pairs as nuclei for magnetization reversal
in elongated particles. As has been reported elsewhere
\cite{prlnucl},  the existence of such nonuniform
nuclei  fact can lead to a substantial
reduction of the coercivity compared to standard theories
of magnetization reversal. Further details of
the statistical mechanical theory of magnetization
reversal are covered in the following paper \cite{braunnucl}.

\section{acknowledgements}

I kindly acknowledge illuminating
discussions with W. Baltensperger,  O. Brodbeck,  J.S. Broz,
J. Ho{\l}yst, S. Skourtis and H. Suhl. This work has been  supported
by the Swiss National Science Foundation and  by ONR-Grant
N00014-90-J-1202.

\appendix
\section*{}
In (\ref{Hk}), (\ref{Hfbar}), (\ref{Hpbar})
we encountered Schr\"odinger operators of the form
\begin{equation}
{\cal H}^{(m)}=-{d^2\over d x^2} -{m(m+1)\over \cosh^2x},
\label{hrefl}
\end{equation}
with $m$ an integer. In the following, we shall show
how the corresponding eigenvalue problems may be
solved exactly with the method of Ref. \onlinecite{infeld}.
The continuum eigenfunctions of ${\cal H}^{(m)}$
have the remarkable property that their reflection
coefficient is zero. We write the eigenvalue problem of
(\ref{hrefl}) as follows
\begin{equation}
{\cal H}^{(m)}\psi^{(m)}_\lambda= \lambda\psi^{(m)}_\lambda.
\label{hev}
\end{equation}
The key point for the solution of the eigenvalue
problem (\ref{hev})
is the observation that ${\cal H}^{(m)}$ may be factorized in
two different ways:
\begin{eqnarray}
{\cal H}^{(m)}&=&{\cal Q}_+^{(m)}{\cal Q}_-^{(m)}-m^2,\label{qqminus}\\
&=&{\cal Q}_-^{(m+1)}{\cal Q}_+^{(m+1)} -(m+1)^2,\label{qqplus}
\end{eqnarray}
with
\begin{equation}
{\cal Q}_\pm^{(m)}=\mp{d\over dx} +m\tanh x.
\label{qdef}
\end{equation}

Operating on (\ref{hev}) from the left with ${\cal Q}_-^{(m)}$,
${\cal Q}_+^{(m+1)}$, we recognize that if $\psi_\lambda^{(m)}$
is an eigenfunction of ${\cal H}^{(m)}$ with eigenvalue
$\lambda$, then
\begin{eqnarray}
\psi_\lambda^{(m-1)}&=&{\cal Q}_-^{(m)} \psi_\lambda^{(m)},\label{psimm1}\\
\psi_\lambda^{(m+1)}&=&{\cal Q}_+^{(m+1)}\psi_\lambda^{(m)},\label{psimm}
\end{eqnarray}
are eigenfunctions of ${\cal H}^{(m-1)}$, ${\cal H}^{(m+1)}$,
respectively, to the {\it same} eigenvalue $\lambda$.

The {\it continuum eigenfunctions} of ${\cal H}^{(m)}$ can
thus be related to those of ${\cal H}^{(m-1)}$.
Since ${\cal H}^{(0)}$ represents the free
problem, the continuum eigenfunctions of ${\cal H}^{(m)}$ can be
obtained by successive application of ${\cal Q}_+^{(m)}$
onto plane wave solutions, i.e.
\begin{equation}
\psi_k^{(m)}={\cal Q}_+^{(m)}\dots {\cal Q}_+^{(2)} {\cal Q}_+^{(1)}
e^{ikx},
\label{cont}
\end{equation}
and belong to the eigenvalue $\lambda=k^2$.
For ${\cal H}^{(1)}=-d^2/dx^2-2{\mathop{\rm sech}\nolimits}^2 x$,
Eq. (\ref{cont}) yields
\begin{equation}
\psi_k^{(1)}={\cal Q}^{(1)}_+ e^{ikx}=[-ik+\tanh x] e^{ikx}
\label{H1cont}
\end{equation}
For ${\cal H}^{(2)}=-d^2/dx^2-6{\mathop{\rm sech}\nolimits}^2 x$ we
obtain the continuum eigenfunctions
\begin{eqnarray}
\psi_k^{(2)}&=&{\cal Q}^{(2)}_+{\cal Q}^{(1)}_+e^{ikx},\\
            &=&[3 \tanh^2 x -3ik\tanh x -1 -k^2] e^{ikx}.
\label{H2cont}
\end{eqnarray}

To find the {\it bound states} with $\lambda<0$, we first remark
that the normalization of bound state eigenfunctions
with different $m$ are related as
\begin{eqnarray}
\int_{-\infty}^{+\infty} dx (\psi_\lambda^{(m-1)})^2
&=&\int_{-\infty}^{+\infty}
dx \psi_\lambda^{(m)} {\cal Q}_+^{(m)}
{\cal Q}_-^{(m)} \psi_\lambda^{(m)},
\nonumber \\
&=&\left(\lambda+m^2\right)\int_{-\infty}^{+\infty} dx
(\psi_\lambda^{(m)})^2.\label{norm}
\end{eqnarray}
Continuing this recursion towards lower values of $m$
we recognize that the normalization of, say,
$\psi_\lambda^{(l-1)}$ would become negative.
This can only be avoided if the recursion (\ref{norm})
stops, i.e. if the bound state eigenvalues are given by
\begin{equation}
\lambda_l=-l^2,\qquad l=1,2,\dots m.
\end{equation}
According to (\ref{norm}) this implies that
\begin{equation}
{\cal Q}_-^{(l)}\psi_{\lambda_l}^{(l)}=0.
\label{break}
\end{equation}
This differential equation can be integrated with (\ref{qdef})
\begin{equation}
\psi_{\lambda_l}(x)={\mathop{\rm sech}\nolimits}^l x.
\end{equation}
For $m>l$, the unnormalized l$^{th}$ bound state (counted from
the continuum) can be obtained recursively
with the help of (\ref{psimm})
\begin{equation}
\psi^{(m)}_\lambda={\cal Q}^{(m)}_+\dots {\cal Q}^{(l+1)}_+
{\mathop{\rm sech}\nolimits}^l x.
\end{equation}
Specifically, we obtain for $m=1$
\begin{equation}
\psi_1^{(1)}={\mathop{\rm sech}\nolimits} x,
\end{equation}
with energy $\lambda_1=-1$, and for $m=2$
\begin{eqnarray}
\psi_2^{(2)}={\mathop{\rm sech}\nolimits}^2 x,\\
\psi_1^{(2)}={\mathop{\rm sech}\nolimits} x \tanh x.
\end{eqnarray}
with energies $\lambda_2=-4$ and $\lambda_1=-1$.
All operators (\ref{hrefl}) share the property of having a zero energy
resonance. This means that an infinitesimal increase in the potential
strength of (\ref{hrefl}) leads to an additional bound state. Therefore, the
occurence of the operators (\ref{hrefl}) as describing fluctuations around
nonlinear structures in some limit of  the  external field has to be
handled with care, since their number of bound states in general differs
from those of the operators they emerge from.

\begin{figure}
\caption{ a) The $\pi$-Bloch wall interpolates between
different anisotropy minima;
b) Fluctuations $\varphi$, $p$ around  a given structure
with  $\phi_{s,\varphi}$ and $\theta=\pi/2$
at a given space point $z$.
\label{blochwall}}
\end{figure}

\begin{figure}
\caption{ The nucleus is shown for a) small fields ($R_s=3.5$) and b)
for fields close to the anisotropy field ($R_s=0.4$)
\label{nucleus}}
\end{figure}

\begin{figure}
\caption{ The $2\pi$-Bloch wall is shown for a) small fields ($R_b=3.5$)
and b) for large fields ($R_b=0.4$)
\label{2pblochwall}}
\end{figure}

\begin{figure}
\caption{Stability and instability regions of the $2\pi$-Bloch wall as a
function of the external field $h$ and the demagnetizing field strength
$Q^{-1}$.
The dotted and dashed lines refer to (\protect\ref{blochinstsmall})
and (\protect\ref{blochinstlarge}) respectively.
\label{instab}}
\end{figure}

\begin{figure}
\caption {The rescaled bound state energies of  ${\cal H}^{s\varphi}$ and
${\cal H}^{s p}$ are shown as a function of
$R_s$ ($\delta_s=\coth R_s$).  The shaded region indicates
the continuum states.
The horizontal lines $E^{s\varphi}_1$ and $E^{sp}_0$ correspond to the zero
modes of ${\cal H}^{s\varphi}$ and ${\cal H}^{b p}$,
respectively. $E^{s\varphi}_2$ is a very weakly bound state
just below the continuum threshold. The dashed lines
indicate the approximation
formulas (\protect\ref{Ef0pert0}) and (\protect\ref{Ep1pert0}).
The bound state energies of
${\cal H}^{b \varphi}$ and ${\cal H}^{b p}$ may be
obtained from (\protect\ref{Eplink}),(\protect\ref{Ebplink}).
\label{boundstates}}
\end{figure}

\begin{figure}
\caption {The  odd and even parity scattering phases of
${\cal H}^{s\varphi}$  are shown for different values of
$R_s$ ($\delta_s=\coth R_s$).
$\bar\Delta^{s\varphi}(k)$ and $\hat\Delta^s(k)$ are given by
(\protect\ref{Dhat}) -
 (\protect\ref{Dpbar}).
Note that the convergence
$\Delta_{(e)}^{s\varphi}(k)\to\bar\Delta^{s\varphi}(k)$ for $R_s\to 0$
and $\Delta_{(e)}^{s\varphi}(k)\to\hat \Delta^{s}(k)$ for $R_s\to\infty$ is
nonuniform.
\label{Hfphases}}
\end{figure}

\begin{figure}
\caption {The  odd and even parity scattering phases of
${\cal H}^{s p}$  are shown .
The convergence $\Delta_{(o)}^{sp}(k)\to\bar\Delta^{sp}(k)$ for
$R_s\to 0$
and $\Delta_{(e)}^{sp}(k)\to\hat \Delta^s(k)$ for $R_s\to\infty$ is
nonuniform.
\label{Hpphases}}
\end{figure}

\begin{figure}
\caption{
Magnetization configurations in  films that can be described
by the model (\protect\ref{e}).
Note that the anisotropy constant has to be chosen as indicated to
incorporate the local part of the magnetostatic interaction.
Note also that the orientation of the coordinate frame in a) as used in
the text is different from  b) and c).
\label{thinfilms}}
\end{figure}

\begin{figure}
\caption{
Domain wall separations for a twisted domain wall pair in
a) perpendicularly  and b) in-plane magnetized  films
due to balance between  external field and  exchange (solid line, Eq.
(\protect\ref{wallsep}) )
or between external field and demagnetizing effects (broken line,
a) Eq. (\protect\ref{widthperp})),   b) Eq. (\protect\ref{widthinplane}).
\label{balance}}
\end{figure}


\begin{references}
\bibitem{magcharge}  Neglecting magnetostatic
surface charges corresponds to an infinite
sample where the limits in transversal directions
are performed before the one along the magnetic chain.
\bibitem{continuity}  The branch of $\arctan$
is always chosen such that $\Delta(k\to\infty)=0$ and
such that $\Delta(k)$ is a continuous function.
\bibitem{demag} This can be seen with the use of the inequality
              $|M_x(z) M_x(z')-M_z(z) M_z(z')|$ $\leq 2 M_0^2$
              and subsequent integration.
\end{references}
\end{document}